\documentclass[aps,preprint]{revtex4}%
\usepackage{amsfonts}
\usepackage{amsmath}
\usepackage{amssymb}
\usepackage{graphicx}%
\begin{document}
\title[Avalanches and Self-Organized Criticality in Superconductors]{Avalanches and Self-Organized Criticality in Superconductors}
\author{Rinke J. Wijngaarden, Marco S. Welling, Christof M. Aegerter and Mariela Menghini}
\affiliation{Department of Physics and Astronomy, Faculty of Sciences, Free University, De
Boelelaan 1081, 1081 HV Amsterdam, The Netherlands}

\begin{abstract}
{We review the use of superconductors as a playground for the experimental
study of front roughening and avalanches. Using the magneto-optical technique,
the spatial distribution of the vortex density in the sample is monitored as a
function of time. The roughness and growth exponents corresponding to the
vortex `landscape' are determined and compared to the exponents that
characterize the avalanches in the framework of Self-Organized Criticality.
For those situations where a thermo-magnetic instability arises, an analytical
non-linear and non-local model is discussed, which is found to be consistent
to great detail with the experimental results. On anisotropic substrates, the
anisotropy regularizes the avalanches.}

\end{abstract}
\date{\today}
\maketitle

\section{Introduction}

\begin{figure}
[t]
\begin{center}
\includegraphics[
width=6.3087cm
]%
{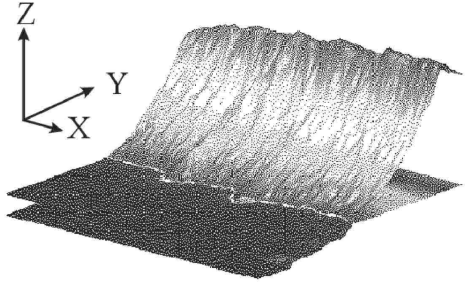}%
\caption{Vortex density (along the vertical axis) in a thin YBa$_{2}$Cu$_{3}%
$O$_{7}$- superconductor as a function of position. Note the similarity with a
pile of sand, including the surface roughness.}%
\label{heap}%
\end{center}
\end{figure}

In this review we discuss the penetration of magnetic flux in thin type-II
superconducting slabs subject to an increasing magnetic field oriented normal
to the slab. In type-II superconductors, magnetic flux penetrates in the form
of vortices, which behave as repulsive particles moving in the plane of the
slab. The vortices can only enter (or leave) at the edge of the slab. Their
motion, however, is hindered by pinning (due to disorder in the sample), which
leads to a gradient in vortex density, somewhat similar to the slope of a sand
pile, see Fig. \ref{heap}. This analogy was already noted a long time ago by
de Gennes\cite{DeGennes1950}. In the present paper, however, this analogy is
extended to the surface roughness of the pile and to the avalanches occurring
on its surface.%

The growth and roughness exponents of this rough surface are determined. Also
the properties of the avalanches occurring on the surface are measured and it
will be shown that they have a power law size distribution function and obey
"finite size scaling". In addition, we find that exponent scaling relations of
Paczuski, Maslov and Bak\cite{PMB1996} are obeyed to reasonable accuracy. All
these observations together make a strong case for the vortex pile system to
belong to the class of Self-Organized Criticality (SOC)\cite{BakHNW} systems.
To investigate the role of disorder, we also study thin films of Niobium,
which are slowly subjected to a hydrogen atmosphere. Hydrogen is absorbed by
Niobium and locally destroys the superconductivity, thus increasing the amount
of disorder. We find that a minimum amount of disorder is necessary for SOC to
occur. In other superconductors, the properties of huge thermo-magnetic
avalanches are studied. These are described very nicely by a recently
developed model\cite{Aranson2005} and can be simulated to high degree of correspondence.

\section{Experimental technique}

Images of the 2-dimensional (i.e. in the plane of the thin film sample)
density of vortices are made using a magneto-optical
technique\cite{KoblWijngMORev}. On top of the sample and in immediate contact
with it, a mirror layer and (directly above this) a 4 $%
\operatorname{\mu m}%
$ thick Yttrium Iron Garnet film with large Faraday effect are placed. The
sample assembly is placed on a custom built insert with $X,Y,Z,\theta
,\phi,\eta$-controls for proper alignment, which fits in a commercial $1%
\operatorname{T}%
$ or $7%
\operatorname{T}%
$ Oxford instruments cryostat. The insert is equipped with lenses, a
semi-transparent mirror and polarizers, making it effectively a polarization
microscope. Where ever there is a local magnetic field in the sample, the
incoming polarization vector is turned, thereby increasing the detected
intensity. Thus the intensity in the camera yields directly the magnetic field
and hence vortex density for each position in the sample. Using a modulation
scheme\cite{RWMOLI,RWOystese} this intensity map becomes linear in field and
much more sensitive.

\section{Roughening of the vortex landscape}

As mentioned above, the vortex density map of the sample is generally not
smooth (see Fig. \ref{heap}) but displays some roughness. For example, one may
define a flux front as the `foot' of the vortex `mountain' as indicated by the
white line in Fig. \ref{heap}. The exact position is determined by the cross
section between the vortex `mountain' and a horizontal plane at a level of 3
times the noise in the vortex-free region. Generally\cite{BaraStan}, for such
fronts, the r.m.s. width $w$ growths initially with time as $w\sim t^{\beta}$.
After some time $t_{\times}$, however, lateral correlations extend over the
whole area of the pile and $w$ does not increase anymore: it is limited to a
pile-size dependent value $w\sim L^{\alpha}$ where $L$ is the linear size of
the pile and $\alpha$ is the roughness exponent. At $t_{\times}$, both
relations hold from which%
\begin{equation}
t_{\times}\sim L^{\alpha/\beta} \label{eq_tcross}%
\end{equation}

To find $\alpha$ and $\beta$ to higher accuracy, we use the correlation
function\cite{BaraStan}%
\begin{equation}
C\left(  x,t\right)  =\left\{  \left\langle \left[  \delta h\left(  \xi
,\tau\right)  -\delta h\left(  \xi+x,\tau+t\right)  \right]  ^{2}\right\rangle
_{\xi,\tau}\right\}  ^{1/2}%
\end{equation}

where $\delta h=h-\overline{h}$ and $h\left(  x,t\right)  $ is the height of
the interface at position $x$ and time $t$, where $\overline{h}=\left\langle
h\right\rangle _{x\in\left[  0,L\right]  }$ and $\left\langle \cdot
\right\rangle _{\eta}$ denotes an average over $\eta$. From this correlation
function, one determines the growth exponent $\beta$ and roughness exponent
$\alpha$ using\cite{BaraStan} $C\left(  x,0\right)  \sim x^{\alpha}$ and
$C\left(  0,t\right)  \sim t^{\beta}$.%
\begin{figure}
[t]
\begin{center}
\includegraphics[
width=7.1171cm
]%
{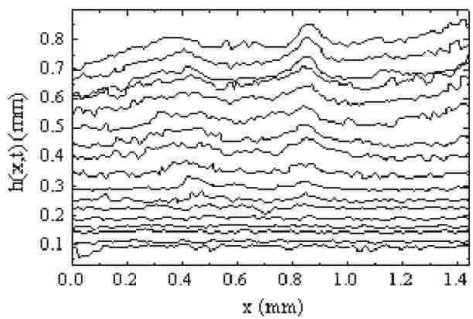}%
\caption{Flux fronts in a YBa$_{2}$Cu$_{3}$O$_{7}$-sample of 80
$\operatorname{nm}$ thick that was deposited on a NdGaO$_{3}$ substrate for
different applied external fields. The first front (bottom) is recorded at 1
mT and subsequent fronts are at 1 mT intervals.}%
\label{YBCOfronts}%
\end{center}
\end{figure}

For YBa$_{2}$Cu$_{3}$O$_{7}$-samples of 80 $%
\operatorname{nm}%
$ thick that were \ deposited on NdGaO$_{3}$ substrates, we recently found for
the roughness of the \textit{2-dimensional surface} of the
pile\cite{WellingEPL2004},\cite{AegerterPA2005}:%
\begin{align}
\alpha &  =0.8\left(  1\right) \label{roughexp}\\
\beta &  =0.6\left(  1\right) \nonumber
\end{align}

The values of these exponents are consistent with the Q-EW (quenched-disorder
Edwards--Wilkinson) model. Apparently the vortices move on a substrate with
randomly distributed quenched pinning sites, as is indeed expected for a
high-T$_{c}$ superconductor. Similar values for $\alpha$ and $\beta$ for the
\textit{flux fronts} in YBa$_{2}$Cu$_{3}$O$_{7}$ thin films were previously
obtained by Surdeanu \textit{et al.}\cite{SurdeanuPRL1999} and for $\alpha$ in
Niobium thin films by Vlasko-Vlasov \textit{et al.}\cite{VlaskoVlasov2004}. In
general, the values of the exponents for the 2-dimensional surface are lower
than for the fronts. The reason for obtaining similar values for $\alpha$ and
$\beta$ for the 1-d and 2-d case is the higher number of defects in the sample
used for 2-d case.

\section{Small vortex avalanches}

\begin{figure}
[t]
\begin{center}
\includegraphics[
width=7.2884cm
]%
{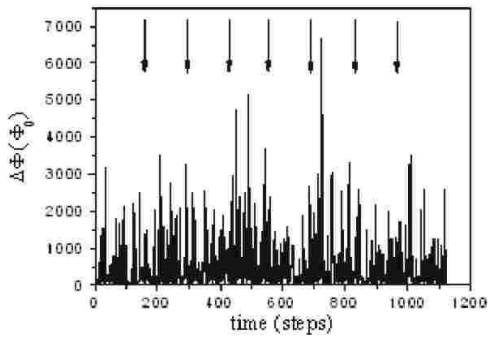}%
\caption{Size (expressed in units of the flux quantum $\Phi_{0}$) of vortex
avalanches in a YBa$_{2}$Cu$_{3}$O$_{7}$-sample of 80 $\operatorname{nm}$
thick that was deposited on a NdGaO$_{3}$ substrate as a function of time for
8 different experiments (separated by the vertical arrows).}%
\label{vortexavals}%
\end{center}
\end{figure}

From a closer inspection of the growth process of the vortex `mountain', shown
on Fig. \ref{heap}, it is clear that the formation of this `mountain' and
hence the flux penetration process is not continuous, but takes place in the
form of avalanches. In our experimental procedure, the external field is
increased in steps of $50%
\operatorname{\mu T}%
$ and after each such increase an image of the vortex density in the sample is
obtained. The size of each avalanche is calculated from
\begin{equation}
s=\frac{1}{2}\int\left\vert \Delta B_{z}\left(  \mathbf{r}\right)  \right\vert
d\mathbf{r}%
\end{equation}

where $\Delta B_{z}\left(  \mathbf{r}\right)  $ is the difference between two
consecutive images \textit{minus} the $50%
\operatorname{\mu T}%
$ increase of the external field. The resulting time series of avalanches is
shown in Fig. \ref{vortexavals}.%

\begin{figure}
[t]
\begin{center}
\includegraphics[
width=6.1835cm
]%
{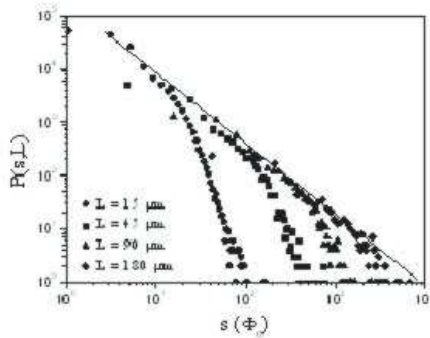}%
\caption{Number $P\left(  s,L\right)  $ of avalanches of size $s$ in a
YBa$_{2}$Cu$_{3}$O$_{7}$-sample of 80 $\operatorname{nm}$ thick that was
deposited on a NdGaO$_{3}$ substrate occuring in a window of observation of
linear dimension $L$. The straigt line indicates the powerlaw in the limit for
large $L$.}%
\label{fsspre}%
\end{center}
\end{figure}

The size distribution of the avalanches is shown for differently sized windows
of observation in Fig. \ref{fsspre} as a log-log plot. From the straight line
behavior in the limit of large systems (\textit{viz.} large windows of
observation), it is clear that the avalanche sizes are power law distributed
with $P\left(  s\right)  \sim s^{-\tau}$, just like earthquakes and many more
natural phenomena\cite{BakHNW}.%

\begin{figure}
[h]
\begin{center}
\includegraphics[
width=6.1022cm
]%
{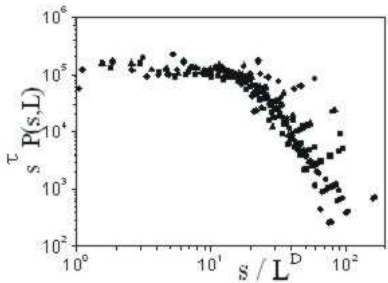}%
\caption{Finite size scaling plot of the data in Fig. \ref{fsspre}.}%
\label{fss}%
\end{center}
\end{figure}

The ubiquitous occurrence of power laws in many natural phenomena lead Bak to
the idea of Self-Organized Criticality\cite{BakHNW}, a model where slowly
driven systems organize themselves automatically to be in a metastable state
where small disturbances can lead to a response of arbitrary size. The
archetypal example is the sand pile with its critical slope (metastability)
and the avalanches of sand sliding down its surface. Within the SOC framework
(and otherwise too), the volume $V$ of an avalanche with fractal dimension $D$
and linear extension $\ell$ is given by $V=\ell^{D}.$ Since the linear size of
the avalanche is limited by the system size $L$ (by neccesity $\ell\lesssim
L$), one expects that the finite size effect-induced deviations from the
straight line in Fig. \ref{fsspre} scale as $L^{D}$. Indeed the finite size
scaling (FSS) plot of Fig. \ref{fss} shows a nice data collapse and
yields\cite{WellingEPL2004} $\tau=1.29\left(  2\right)  $ and $D=1.89\left(
3\right)  $.%

Apart from the FSS and the power law behavior there is another check one can
make for SOC. This is based on the exponent relations of Paczuski, Maslow and
Bak\cite{PMB1996}. We discuss and use two such relations.

The first scaling relation is found by calculating the volume $V$ of the
fractal avalanche. By definition, $V\sim L^{D}$. Also, this volume equals the
fractal surface times the fractal height. The fractal surface area is the
projection of the avalanche cluster on an average (flat) plane $\mathbf{P}$
through the pile surface. This area is $L^{d_{B}}$ where $d_{B}$ is the
surface fractal dimension of the avalanche cluster. The height or thickness of
an avalanche is obtained by subtracting the heights (in the direction
perpendicular to $\mathbf{P}$) of the piles before and after the avalanche.
Since the avalanche modifies the pile only locally, this fractal thickness
scales as the surface roughness and is proportional to $L^{\alpha}$. Hence $V$
$\sim L^{d_{B}}L^{\alpha}$. Combining both expressions for $V$ yields
$L^{D}\sim L^{d_{B}}L^{\alpha}$ from which we obtain the scaling relation%
\begin{equation}
D=d_{B}+\alpha\label{eq_pmbalpha}%
\end{equation}

The second scaling relation follows from the fact that the deviation from
power law behavior seen in Fig. \ref{fsspre} occurs because avalanches above a
certain size `feel' that the pile is finite.

%

\begin{figure}
[h]
\begin{center}
\includegraphics[
width=6.4185cm
]%
{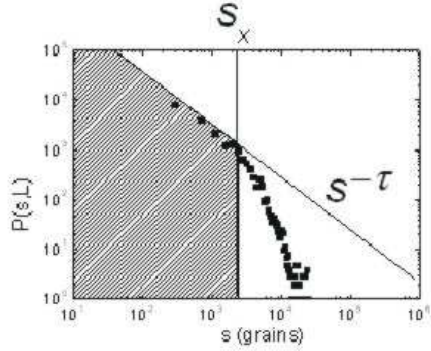}%
\caption{The scaling relation $D\left(  2-\tau\right)  =\frac{\alpha}{\beta}$
is derived by comparing the time needed for correlations to span the whole
pile and the time needed to create avalanches that span the whole pile (see
text).}%
\label{pmbbeta}%
\end{center}
\end{figure}

This, of course, occurs at a time $t_{\times}$ (see eq. \ref{eq_tcross}), when
correlations and avalanches start to span the whole `mountain'. By definition
the size of such an avalanche is $s_{\times}$. Since we increase the external
field at constant rate, $t_{\times}$ also is proportional to the number of
vortices $M$ we must add to the pile to obtain such a `mountain'-spanning
avalanche. However, before we create an avalanche of size $s_{\times}$, many
smaller avalanches have occurred, adding to the number of vortices that we
have to add before a `mountain'-spanning avalanche takes place. Thus $M$ is
equal to the integral of the size distribution function up to $s_{\times}$,
i.e. the shaded area in Figure \ref{pmbbeta}. This gives%
\begin{equation}
t_{\times}\sim M=\int_{0}^{s_{\times}}sP\left(  s\right)  ds\sim\int
_{0}^{s_{\times}}s\;s^{-\tau}ds\sim s_{\times}^{2-\tau} \label{txpile}%
\end{equation}

In the FFS analysis we found that $s\sim L^{D}$, hence we obtain%

\begin{equation}
t_{\times}\sim s_{\times}^{2-\tau}\sim L^{D\left(  2-\tau\right)  }%
\end{equation}

Combination with eq.\ref{eq_tcross} yields the exponent relation%
\begin{equation}
D\left(  2-\tau\right)  =\frac{\alpha}{\beta} \label{eq_pmbbeta}%
\end{equation}

Equations \ref{eq_pmbalpha} and \ref{eq_pmbbeta}, which were previously
derived by Paczuski \textit{et al.}\cite{PMB1996}, offer an interesting
possibility: one can calculate the roughness and growth exponents, $\alpha$
and $\beta$, from the \textit{avalanche} properties only. In the table below
we compare the values from such an analysis with those obtained above from a
direct roughness analysis of the surface of the pile:

$%
\begin{tabular}
[c]{|l|l|l|}\hline
& $\alpha$ & $\beta$\\\hline
from roughness analysis & $0.8\left(  1\right)  $ & $0.6\left(  1\right)
$\\\hline
from avalanche analysis & $0.7\left(  1\right)  $ & $0.5\left(  1\right)
$\\\hline
\end{tabular}
\ $

Clearly, a good agreement is found, which supports the underlying assumption
i.e. that SOC theory yields a valid description of the avalanche behavior in
this superconductor.

\section{Transition from non-SOC to SOC}

\begin{figure}
[t]
\begin{center}
\includegraphics[
width=6.4581cm
]%
{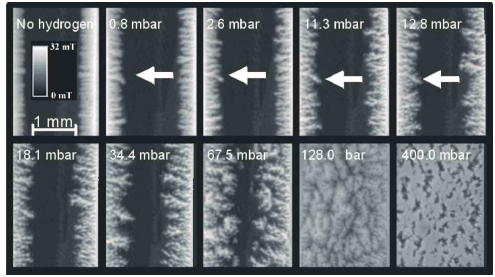}%
\caption{Magnetic flux distribution in a 500 nm thick Nb thin film on an
R-plane sapphire substrate after zero field cooling to 4.2 K and subsequently
applying a field of 6 mT, for different hydrogen loading pressures, as
indicated. The black area corresponds to the vortex-free phase. Bright regions
correspond to the vortex-phase. The arrows indicate a branching magnetic flux
protrusion, of which the branching is increasing with increasing hydrogen
content.}%
\label{NbH}%
\end{center}
\end{figure}

Naturally, one may wonder whether the avalanche and roughness properties in
\textit{all} superconductors are governed by SOC. To investigate this
question, we study these properties as a function of quenched disorder. As a
sample we use a 500 nm thick Niobium thin film on an $R$-plane ($1\bar{1}02$)
sapphire substrate, which is a superconductor with $T_{c}=9.25%
\operatorname{K}%
$, but which also absorbs hydrogen from the surrounding atmosphere. The
hydrogen enters the Niobium lattice interstitially and at low temperatures
precipitates in small non-superconducting clusters, that (like other defects)
constitute strong pinning centers. The hydrogen gas absorption was always done
at room temperature by applying a certain hydrogen gas pressure for $1$ hour.
Typical flux penetration patterns are shown in Fig. \ref{NbH}.%

Avalanches are observed during the slow increase of the externally applied
magnetic field, in similar manner as described above. The avalanches in the
hydrogen-free sample are quite compact and all of similar size, while those in
the hydrogenated sample (e.g. at 18.1 $%
\operatorname{mbar}%
$) are more fractal and have a large range of sizes. As a consequence, a
finite size scaling analysis\cite{WellingNbH2004} \textit{does} work for the
hydrogenated sample, but not for the hydrogen free sample. This, together with
a significant change in fractal dimension\cite{WellingNbH2004}, indicates that
(at least for these samples) SOC occurs only if a minimal amount of static
disorder is present. We conjecture that this is a more general phenomenon and
that also on granular piles SOC behavior depends on the possibility that
disorder (randomness) can develop on its surface.

\section{Avalanches in patterned superconductors}

\begin{figure}
[t]
\begin{center}
\includegraphics[
width=6.2955cm
]%
{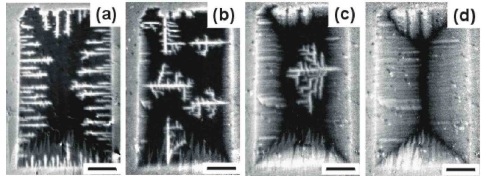}%
\caption{Images of a Pb sample with square array of antidots. Fingerlike
dendritic penetration occurs (a) at $\mu_{0}H=1.2\operatorname{mT}$ and
$T=4.5\operatorname{K}$ and at (b) $\mu_{0}H=1.2\operatorname{mT}$ and
$T=5.5\operatorname{K}$, treelike dendritic outburst coexisting with smooth
flux penetration occurs at (c) $\mu_{0}H=1.5\operatorname{mT}$ and
$T=6.0\operatorname{K}$, and a smooth profile is found at (d) $\mu
_{0}H=1.5\operatorname{mT}$ and $T=6.5\operatorname{K}$. In the bottom part of
the images a saw-tooth-like articact \ from the magneto-optical garnet is
visible. The scale bar in each figure corresponds to 0.5 mm.}%
\label{fig_antidots}%
\end{center}
\end{figure}

If a superconductor is perforated with a square array of tiny holes ($1-10%
\operatorname{\mu m}%
$ in diameter and designated as `anti-dots'), the moving vortices tend to be
guided in the direction of the lattice vector of these
holes\cite{Pannetier2003}. In this case, avalanches are still possible, but
they are guided in similar manner, reducing their fractal dimension, see Fig.
\ref{fig_antidots}. However, still a power law distribution of avalanche sizes
is observed\cite{Menghini2005}: the guiding apparently does not destroy SOC behavior.%

\section{Thermomagnetic Avalanches}

\begin{figure}
[t]
\begin{center}
\includegraphics[
width=6.2867cm
]%
{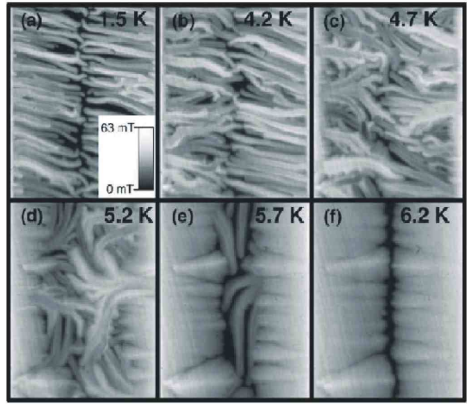}%
\caption{Magnetic flux distribution in Nb thin film on an A-plane sapphire
substrate at 40.0 mT after cooling in zero field to the temperatures
indicated. The scale-bar indicates the local magnetic field.}%
\label{NbA}%
\end{center}
\end{figure}

Interestingly, in very similar samples, 500 nm thick Niobium thin films on an
$A$-plane ($11\bar{2}0$) sapphire substrate (and without anti-dots), a
completely different behavior is observed, see Fig. \ref{NbA}.
At low temperatures ($T\lesssim5.7%
\operatorname{K}%
$), upon increasing the field after cooling in zero field, initially, the
vortex density is rather smooth, but somewhat plume-like, similar to what is
shown in Fig. \ref{NbA}f. After exceeding a certain distance from the edge,
however, large avalanches occur, which span immediately nearly half of the
width of the sample. At low temperature (Fig. \ref{NbA}a) these avalanches are
much more narrow and less branched than at higher temperature (Fig.
\ref{NbA}e). After each avalanche, the `smooth' front starts again from the
sample edge and runs over the previous, still existing, vortex landscape. As
soon as it is close to the previous front, a new avalanche may occur. The
avalanches do not reproduce in the same spot from one experiment to the next,
indicating that the phenomenon is due to an intrinsic instability and not to
defects in the samples. Recently Aranson \textit{et al.}\cite{Aranson2005}
developed a nice theoretical model, which was also used as the basis for
numerical simulations, which very closely resemble the experimental results
(similar experimental results were obtained before, see refs 3-9 of ref.
\cite{Aranson2005}, while similar models were developed by Shantsev \textit{et
al.}\cite{Shantsev2003} and Baggio\textit{\ et al.}\cite{Baggio2005}). In this
model, the system is described by two coupled differential equations:%
\begin{align}
\dot{\theta}  &  =\nabla^{2}\theta-\theta+\gamma\mathbf{j}^{2}%
r\label{thermomagn}\\
\tau\dot{g}  &  =\hat{K}\left\{  \frac{\partial}{\partial x}\left[
r\frac{\partial g}{\partial x}\right]  +\frac{\partial}{\partial y}\left[
r\frac{\partial g}{\partial y}\right]  -\tau\dot{H}_{0}\right\} \nonumber
\end{align}

where $\theta$ is a reduced temperature, $\mathbf{j}$ is the current density
and $r=r\left(  j,\theta\right)  $ the resistivity of the superconductor, $g$
is a local magnetization, defined by $\mathbf{j}=\nabla\times g\left(
\mathbf{s}\right)  \hat{z}$. For thin film samples, like ours, the second
equation is highly non-local, which is described by the kernel $\hat{K}$. The
parameter $\gamma$ contains the thermal properties of the film and its contact
to the substrate, while the parameter $\tau$ is the ratio between magnetic and
thermal diffusion times. Small values of $\tau$ make the system more unstable.
For a detailed comparison between model and experiment see ref.
\cite{AronsonAdditional}.

\section{Conclusion}

We find evidence for SOC behavior of avalanches on the vortex landscape of
many different superconductors, provided that enough (quenched) disorder is
present. This SOC (avalanche) behavior is preserved in the presence of a
square array of strong pins (anti-dots). For superconductors close to thermal
runaway, only large avalanches occur, all of similar size (non-SOC behavior).

\section{Acknowledgment}

We thank I.S. Aranson, A. Gurevich, V.K. Vlasko-Vlasov, V.M. Vinokur, U. Welp,
A. Silhanek, S. Raedts and V. Moshchalkov for fruitful discussions. This work
was supported by FOM (Stichting voor Fundamenteel Onderzoek der Materie),
which is financially supported by NWO (Nederlandse Organisatie voor
Wetenschappelijk Onderzoek). We acknowledge the VORTEX program of the European
Science Foundation for support.

------------------

\end{document}